\documentclass[aps,pra,showpacs,amssymb,twocolumn]{revtex4-1}
\usepackage{graphicx}
\usepackage{appendix}
\usepackage{color}
\usepackage[position=top,singlelinecheck=off,justification=raggedright,caption=false]{subfig}
\usepackage{epstopdf} 
\DeclareGraphicsExtensions{.pdf,.eps,.png,.jpg,.mps}
\usepackage[colorlinks=true,citecolor=blue,urlcolor=blue,linkcolor=blue]{hyperref}
\usepackage{bm,amsmath}
\usepackage{dcolumn}

\newcommand{\bra}[1]{\left\langle #1 \right|}
\newcommand{\ket}[1]{\left|#1\right\rangle}
\newcommand{\braket}[2]{\left\langle#1 |  #2\right\rangle}

\def\BEq{\begin{equation}}
\def\EEq{\end{equation}}
\def\BEqA{\begin{eqnarray}}
\def\EEqA{\end{eqnarray}}
\def\BW{\begin{widetext}}
\def\EW{\end{widetext}}

\begin{document}

\title{Emulating quantum state transfer through a spin-1 chain on a 1D lattice of superconducting qutrits}

\author{Joydip Ghosh}
\email{ghoshj@ucalgary.ca}
\affiliation{Institute for Quantum Science and Technology, University of Calgary, Calgary, Alberta T2N 1N4, Canada}

\date{\today}

\begin{abstract}
Spin-1 systems, in comparison to spin-$\frac{1}{2}$ systems, offer a better security for encoding and transfer of quantum information, primarily due to their larger Hilbert spaces. Superconducting artificial atoms possess multiple energy-levels, thereby capable of emulating higher-spin systems. Here we consider a 1D lattice of nearest-neighbor-coupled superconducting transmon systems, and devise a scheme to transfer an arbitrary qutrit-state (a state encoded in a three-level quantum system) across the chain. We assume adjustable couplings between adjacent transmons, derive an analytic constraint for the control-pulse, and show how to satisfy the constraint to achieve a high-fidelity state-transfer under current experimental conditions. Our protocol thus enables enhanced quantum communication and information processing with promising superconducting qutrits.
\end{abstract}

\pacs{03.67.Ac, 85.25.-j, 05.40.Fb} 

\maketitle

\section{Introduction}
\label{sec:Introduction}
Quantum State Transfer (QST) between two quantum systems remains a primitive operation for many protocols in quantum communication, simulation and information processing. QST along a chain of nearest-neighbor-coupled spin-$\frac{1}{2}$ systems has been extensively studied as a channel for short-distance quantum communication~\cite{PhysRevLett.91.207901,PhysRevLett.93.230502,PhysRevA.69.034304,PhysRevA.89.062301,PhysRevLett.101.230502,doi:10.1080/00107510701342313,doi:10.1142/S0219749910006514}, and its implementations have been proposed for NMR systems~\cite{PhysRevLett.99.250506,PhysRevA.85.030303,1367-2630-14-8-083005}, trapped Rydberg ions~\cite{1367-2630-10-9-093009}, coupled-cavity-arrays~\cite{liu2014transfer} and superconducting flux qubits~\cite{1367-2630-7-1-181}, with experimental realizations reported so far for NMR systems~\cite{rao2013simulation}, photonic lattices~\cite{Bellec:12,PhysRevA.87.012309} and cold atoms~\cite{Fukuhara2013a,Fukuhara2013}. However, with the discovery that quantum information processing becomes more robust on higher-dimensional spin systems~\cite{PhysRevA.61.062308,PhysRevA.67.012311}, considerable attention has been paid to the higher-dimensional spin chains. This leads to the emergence of a number of proposals in recent years for possible QST schemes on $d$-level ($d > 2$) spin chains, specifically on spin-1 chains~\cite{1367-2630-9-5-155,PhysRevA.87.012339,Asoudeh2014QIPraey,PhysRevA.89.062302,PhysRevA.75.050303,wiesniak2013translating,Delgado200722}.

Superconducting artificial atoms contain more than two energy levels that can be readily manipulated and reliably measured, thereby allowing the possibility of emulating the higher spin systems~\cite{Neeley07082009}. In this work, we devise a scheme to emulate a QST along a spin-1 chain on a 1D array of nearest-neighbor-coupled superconducting transmon systems~\cite{Barends2014}. The transmons are treated as \emph{qutrits} (three-level systems) with the three lowest energy levels mapping to the three possible states of a spin-1 particle. We also assume an adjustable coupling between each pair of adjacent transmons that can be tuned via control electronics, an architecture often referred to as a \emph{gmon} device~\cite{chen2014qubit,geller2014tunable}. It should be emphasized in this context that, when two transmons are coupled (via an inductive tunable coupler), the coupling strengths in the single- and double-excitation subspaces are unequal requiring two different timescales to transfer quantum states for those two subspaces. These unequal coupling strengths, in fact, preclude a direct generalization from a qubit-to-qubit state-transfer to a qutrit-to-qutrit state-transfer for superconducting systems, which motivates us to develop a strategy for such a higher-dimensional state-transfer across the chain of superconducting qutrits under experimental conditions.

The problem of emulating the QST on the array of coupled transmon qutrits can be described as follows: First, we prepare an arbitrary qutrit-state $\ket{\psi}=\alpha\ket{0}+\beta\ket{1}+\gamma\ket{2}$ in the first qutrit (as demonstrated by Neeley et al.~\cite{Neeley07082009}), and then control the tunable coupling strengths for a specific time-duration, such that,
\BEq
\label{eq:DefQST}
\ket{\psi}_{1}\otimes\ket{0}_{2}\otimes\ket{0}_{3}\otimes \ldots \otimes\ket{0}_{N} \longrightarrow \ket{0}_{1}\otimes\ket{0}_{2}\otimes\ket{0}_{3}\otimes \ldots \otimes\ket{\psi}_{N},
\EEq 
where the subscripts denote the qutrit-indices and $N$ is the number of transmons in the array. The transformation shown in Eq.(\ref{eq:DefQST}) is achieved via successive state-transfers between adjacent qutrits, given by,
\BEq
\ket{\psi}_{j}\otimes\ket{0}_{j+1} \longrightarrow \ket{0}_{j}\otimes\ket{\psi}_{j+1}, \; {\forall}j\in\{1,2,\ldots,N-1\}.
\EEq
Note that, in order to perform the state-transfer between adjacent qutrits, it is necessary and sufficient that the operations,
\BEq
\begin{array}{l}
\ket{1}_{j}\otimes\ket{0}_{j+1} \longrightarrow \ket{0}_{j}\otimes\ket{1}_{j+1} \\
\ket{2}_{j}\otimes\ket{0}_{j+1} \longrightarrow \ket{0}_{j}\otimes\ket{2}_{j+1},
\end{array}
\EEq
are performed simultaneously with other states unchanged. Here we show how to achieve such a simultaneous state transfer with superconducting qutrits under current experimental constraints.

The remainder of the paper is organized as follows: We first discuss the state transfer between two coupled qutrits in Sec.~\ref{sec:QSTtwoQutrits}. Next, we describe our QST protocol across the array of coupled qutrits in Sec.~\ref{sec:populationTransfer}. The effects of intrinsic and decoherence-induced errors are discussed in Sec.~\ref{sec:errors}, and we conclude with possible future directions in Sec.~\ref{sec:conclusions}.

\section{Quantum state transfer between two qutrits}
\label{sec:QSTtwoQutrits}

Here we focus on the QST between two coupled superconducting qutrits. First we describe the coupled-qutrit model and then discuss our state-transfer protocol.

\subsection{Coupled-qutrit model}
\label{sec:coupledQutrit}

The Hamiltonian of a system of two superconducting transmon devices coupled via an adjustable inductive coupling (the `gmon' architecture~\cite{chen2014qubit,geller2014tunable}) is given by (from the lab-frame),
\BEq
\label{eq:labHamiltonian}
H(t)=\sum_{i=1}^{2}\left[ \begin{array}{ccc}
0 & 0 & 0 \\
0 & \epsilon_{i}(t) & 0 \\
0 & 0 & 2\epsilon_{i}(t)-\eta_{i} \end{array} \right]_{{\rm q}_i}+g(t)X_{\rm 1}X_{\rm 2},
\EEq
where,
\BEq
\label{eq:defX}
X_{k}=\left[ \begin{array}{ccc}
0 & 1 & 0 \\
1 & 0 & \sqrt{2} \\
0 & \sqrt{2} & 0 \end{array} \right]_{{\rm q}_{k}},
\EEq
where $k$ denotes the qutrit index and the matrix subscripts $q_{\rm 1,2}$ denote the matrix representations of the corresponding operators for the ${ \rm 1^{st}}$ and the ${\rm 2 ^{nd}}$ qutrit respectively. $\epsilon_{i}$ in Eq.(\ref{eq:labHamiltonian}) denotes the frequency of the $i^{\rm th}$ qutrit that can be tuned with external control electronics. $g$ denotes the adjustable coupling strength between two qutrits that can be varied between $0$ and $55$ MHz \cite{chen2014qubit}. $\eta_{i}$ is the anharmonicity of the $i^{\rm th}$ qutrit, and here we assume $\eta_{\rm 1}=\eta_{\rm 2}=\eta$ (= 200 MHz) \cite{PhysRevA.87.022309}.

In order to transform our Hamiltonian (\ref{eq:labHamiltonian}) from lab frame to a rotating frame, we specify a local reference clock for each qutrit (with frequencies $\omega_{\rm 1}$ and $\omega_{\rm 2}$) with a clock Hamiltonian,
\BEq
\label{eq:clockHamiltonian}
H_{\rm cl}=\left[ \begin{array}{ccc}
0 & 0 & 0 \\
0 & \omega_{\rm 1} & 0 \\
0 & 0 & 2\omega_{\rm 1} \end{array} \right]_{\rm q_{\rm 1}}+\left[ \begin{array}{ccc}
0 & 0 & 0 \\
0 & \omega_{\rm 2} & 0 \\
0 & 0 & 2\omega_{\rm 2} \end{array} \right]_{\rm q_{\rm 2}}.
\EEq
The unitary operator corresponding to the rotating frame specified by the clock-Hamiltonian (\ref{eq:clockHamiltonian}) is defined as,
\BEq
R(t) \equiv e^{iH_{\rm cl}t}.
\EEq
The Hamiltonian from the rotating frame is then given by,
\BEqA
\label{eq:rotatingHamiltonian}
\widetilde{H}(t) &=& R^{\dagger}(t)H(t)R(t)-i\dot{R^{\dagger}}(t)R(t) \nonumber \\
&=& \sum_{i=1}^{2}\left[ \begin{array}{ccc}
0 & 0 & 0 \\
0 & \Delta_{i}(t) & 0 \\
0 & 0 & 2\Delta_{i}(t)-\eta_{i} \end{array} \right]_{\rm q_{i}}+g(t)V,
\EEqA
where,
\BEqA
&&\Delta_{\rm 1,2}(t) = \epsilon_{\rm 1,2}(t)-\omega_{\rm 1,2}, \nonumber \\ \nonumber \\
&& V = \left[ \begin{array}{ccc}
0 & A & 0 \\
B & 0 & A\sqrt{2} \\
0 & B\sqrt{2} & 0 \end{array} \right],  \;\;\; {\rm with} \;\; \nonumber \\
&& A := \left[ \begin{array}{ccc}
0 & e^{i(\omega_{\rm 1}+\omega_{\rm 2})} & 0 \\
e^{i(\omega_{\rm 1}-\omega_{\rm 2})} & 0 & \sqrt{2}e^{i(\omega_{\rm 1}+\omega_{\rm 2})} \\
0 & \sqrt{2}e^{i(\omega_{\rm 1}-\omega_{\rm 2})} & 0 \end{array} \right], \nonumber \\ \nonumber \\
&& B := \left[ \begin{array}{ccc}
0 & e^{i(\omega_{\rm 2}-\omega_{\rm 1})} & 0 \\
e^{-i(\omega_{\rm 1}+\omega_{\rm 2})} & 0 & \sqrt{2}e^{i(\omega_{\rm 2}-\omega_{\rm 1})} \\
0 & \sqrt{2}e^{-i(\omega_{\rm 1}+\omega_{\rm 2})} & 0 \end{array} \right].
\EEqA
Note that the interaction term $V$ in Eq.(\ref{eq:rotatingHamiltonian}) contains rapidly oscillating elements rotating with a frequency $\omega_{\rm 1}+\omega_{\rm 2}$. Assuming $\omega_{\rm 1}=\omega_{\rm 2}$ (a global clock) and applying Rotating Wave Approximation (RWA) remove these rapidly oscillating terms, for which the Hamiltonian (\ref{eq:rotatingHamiltonian}) can be expressed as,
\BEq
\label{eq:Htilde}
\small{
\widetilde{H}(t)=\sum_{i=1}^{2}\left[ \begin{array}{ccc}
0 & 0 & 0 \\
0 & \Delta_{i}(t) & 0 \\
0 & 0 & 2\Delta_{i}(t)-\eta_{i} \end{array} \right]_{{\rm q}_{i}}}+\frac{g(t)}{2}\left(X_{1}X_{2}+Y_{1}Y_{2}\right),
\EEq 
where,
\BEq
\label{eq:defY}
Y_{k}=\left[\begin{array}{ccc}
0 & -i & 0 \\
i & 0 & -i\sqrt{2} \\
0 & i\sqrt{2} & 0 \end{array} \right]_{{\rm q}_k}, \;\; \forall k \in \{1,2\},
\EEq
and $X_{k}$ is defined in Eq.(\ref{eq:defX}). $\Delta_{\rm 1,2}$ are time-dependent frequencies of the qutrits from the rotating frame that can be varied within $-2.5$ to $+2.5$ GHz using control electronics. Also, it is interesting to note that the transformation from lab-frame to rotating frame, in fact, changes the interaction part of our Hamiltonian from `XX' type to `XY' type under RWA.

\subsection{Population transfer between two qutrits}
Now we describe how to transfer the population from one qutrit to another. In order to perform the population transfer, it is sufficient to transform $\ket{00} \leftrightarrow \ket{00}$, $\ket{10} \leftrightarrow \ket{01}$, and $\ket{20} \leftrightarrow \ket{02}$ simultaneously. These simultaneous transformations can be achieved by bringing the qutrits in resonance (i.e., $\Delta_{\rm 1}$=$\Delta_{\rm 2}$) and then turning the coupling on under certain constraints that we derive analytically in this section.

First, it is important to note that the $\ket{00}$ state is sufficiently detuned from all other energy levels when the qutrits are in resonance, and therefore remains invariant even if the coupling is turned on. We represent the Hamiltonian (\ref{eq:Htilde}) in the single-excitation subspace $\{\ket{01},\ket{10}\}$ (denoted by $\widetilde{H}_{\rm 1}$) and double-excitation subspace $\{\ket{11},\ket{02},\ket{20}\}$ (denoted by $\widetilde{H}_{\rm 2}$) as (after energy rescaling and with $\Delta_{\rm 1}=\Delta_{\rm 2}$),
\BEq
\widetilde{H}_{\rm 1}(t)=\left[\begin{array}{cc}
0 & g \\
g & 0 \end{array} \right] \;\; {\rm and} \;\;
\widetilde{H}_{\rm 2}(t)=\left[\begin{array}{ccc}
\eta & g\sqrt{2} & g\sqrt{2} \\
g\sqrt{2} & 0 & 0 \\
g\sqrt{2} & 0 & 0 \end{array} \right],
\EEq
where the time-dependence is embedded in $g$. In the notation of Pauli spin matrices, $\widetilde{H}_{\rm 1}(t)=g(t)\sigma^{x}$, and therefore, a population transfer in the single excitation subspace requires,
\BEq
\int\limits_{0}^{t_{\rm QST}}g(t)dt=\frac{m\pi}{2},
\label{eq:SWAPatSES}
\EEq
where $m$ is an odd number and $t_{\rm QST}$ denotes the time required for the quantum state transfer.

How about a population transfer in the $\{\ket{02},\ket{20}\}$ subspace ? Note that, the levels $\ket{02}$ and $\ket{20}$ are not directly coupled, but coupled via $\ket{11}$ state. The instantaneous eigenvalues of $\widetilde{H}_{\rm 2}$ are $0$ and $\eta/2\pm\sqrt{(\eta/2)^{2}+(2g)^{2}}$, when the qutrits are in resonance. We can, therefore, construct an effective coupling $g_{\rm eff}$ between $\ket{02}$ and $\ket{20}$ states from the level repulsion between these states, which is given by,
\BEq
g_{\rm eff}=\left|\frac{\eta}{4}-\sqrt{\left(\frac{\eta}{4}\right)^{2}+g^{2}}\right|.
\EEq
Following the same argument as for single excitation subspace, we can express the condition for population transfer between $\ket{20}$ and $\ket{02}$ states as, 
\BEq
\int\limits_{0}^{t_{\rm QST}}g_{\rm eff}(t)dt=\int\limits_{0}^{t_{\rm QST}}\left|\frac{\eta}{4}-\sqrt{\left(\frac{\eta}{4}\right)^{2}+[g(t)]^{2}}\right|dt=\frac{l\pi}{2},
\label{eq:SWAPatDES}
\EEq
where $l$ is an odd number. Since $g \gg g_{\rm eff}\;(\approx 2g^{2}/\eta)$ (assuming $\eta \gg g$), the population transfer in the single excitation subspace is faster than that in the double excitation subspace, which motivates us to assume $l=1$ and $m > 1$. Now, combining Eq.(\ref{eq:SWAPatSES}) and Eq.(\ref{eq:SWAPatDES}), we obtain the condition for population transfer between qutrits as,
\BEq
\int\limits_{0}^{t_{\rm QST}}g(t)dt=m\int\limits_{0}^{t_{\rm QST}}\left|\frac{\eta}{4}-\sqrt{\left(\frac{\eta}{4}\right)^{2}+[g(t)]^{2}}\right|dt=\frac{m\pi}{2}\;,
\label{eq:SWAPqutrit}
\EEq
where $m$ is an odd number and we later show that it is possible to constrain $g$ within an experimentally feasible range for $m=3$.

\subsection{Designing a control-pulse for $g(t)$}

\begin{figure}[htb]
\centering
\subfloat[]{
\label{fig:gPulse}
\includegraphics[angle=0,width=0.9\linewidth]{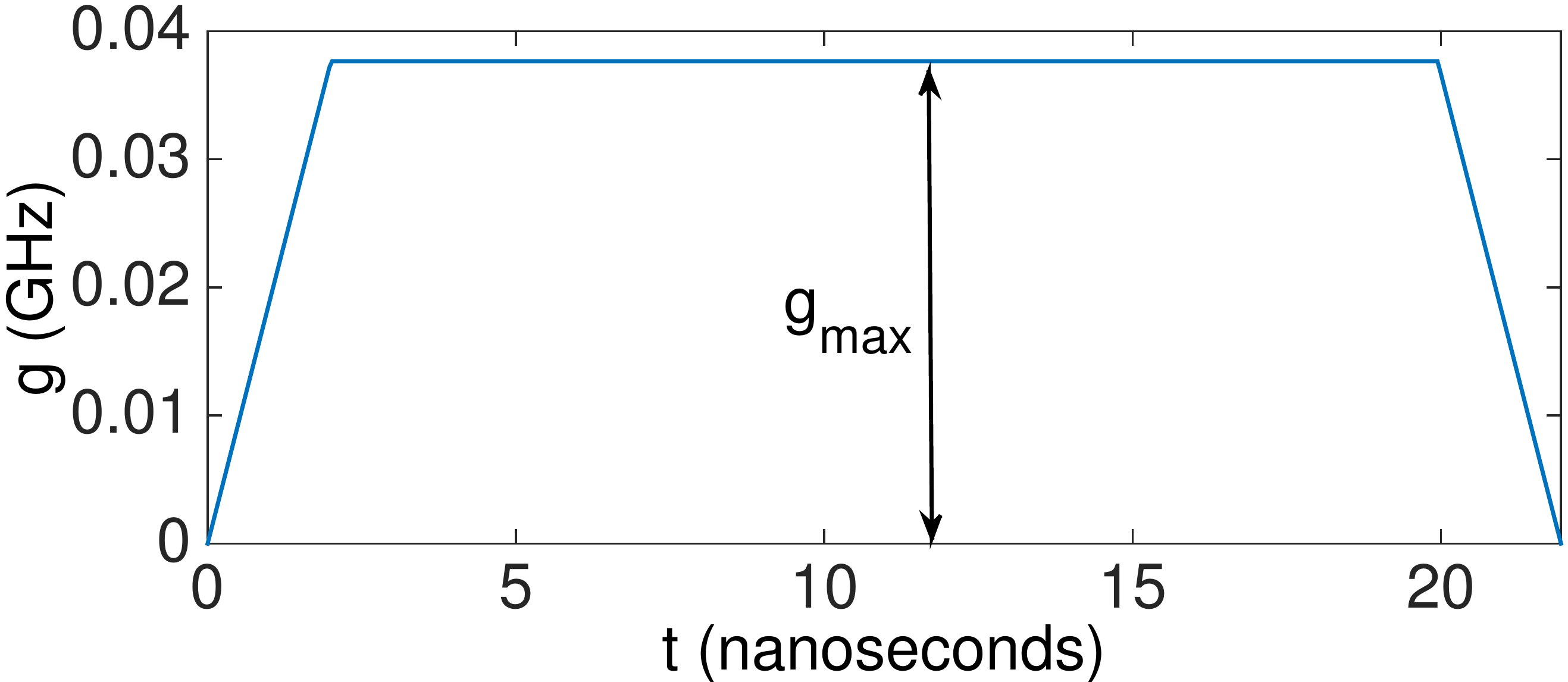}
}
\\
\subfloat[]{
\label{fig:stateTransfer}
\includegraphics[angle=0,width=0.9\linewidth]{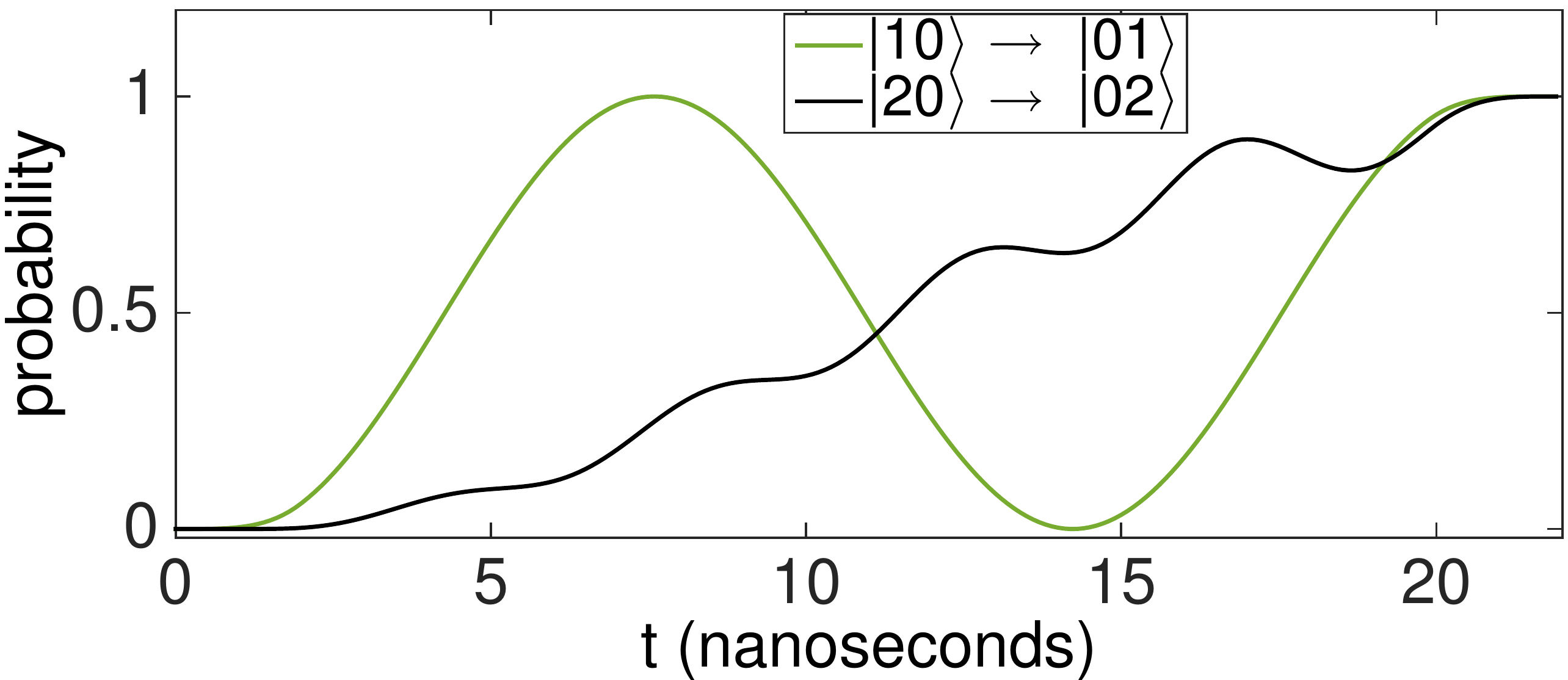}
}
\caption{(Color online) (a) Optimal trapezoidal control-pulse for $g(t)$ while two qutrits are in resonance. (b) Probability of population in the $\ket{01}$ and $\ket{02}$ states under the trapezoidal pulse, assuming that the $\ket{10}$ and $\ket{20}$ states are occupied initially.}
\end{figure}

Now we use Eq.(\ref{eq:SWAPqutrit}) to design a trapezoidal pulse for $g(t)$ with $g(0)=g(t_{\rm QST})=0$ \footnote{One can construct an arbitrary pulse shape that satisfies our Eq.(\ref{eq:SWAPqutrit}). However, we here analyze trapezoidal pulse as it is analytically tractable as well as closely approximates a realistic pulse generated by the control electronics for superconducting circuits}. Let $g_{\rm max}$ be the maximum value that $g(t)$ achieves in the intermediate time, which gives,
\BEq 
\int\limits_{0}^{t_{\rm QST}}g(t)dt=g_{\rm max}(t_{\rm QST}-2)=\frac{3\pi}{2},
\label{eq:tQST}
\EEq
assuming $m=3$ and a $2$ ns ramp as shown in Fig.~\ref{fig:gPulse}. The 2-ns ramp is consistent with the bandwidth specification of existing superconducting control electronics \cite{chen2014qubit}. 

Now, we estimate an approximate value for $g_{\rm max}$, assuming that the area traced out by $g(t)$ and $g_{\rm eff}(t)$ during the constant part of the trapezoidal pulse are almost equal, which essentially means,
\BEq
g_{\rm max}=3\left|\frac{\eta}{4}-\sqrt{\left(\frac{\eta}{4}\right)^{2}+g_{\rm max}^{2}}\right|.
\label{eq:gmax}
\EEq 
Solving for $g_{\rm max}$ from Eq.(\ref{eq:gmax}) and then $t_{\rm QST}$ from Eq.(\ref{eq:tQST}), we obtain,
\BEq
\label{eq:theoreticalParameters}
g_{\rm max}=\frac{3\eta}{16}\;\;\;\;{\rm and}\;\;\;t_{\rm QST}=2+\frac{8\pi}{\eta}.
\EEq
For $\eta=200$ MHz, $g_{\rm max}=37.5$ MHz and $t_{\rm QST}=22$ ns.

It is possible to further improve the performance of qutrit-qutrit population transfer by optimizing $g_{\rm max}$ and $t_{\rm QST}$ independently, using the analytical values as initial solutions. Fig.~\ref{fig:gPulse} shows such an optimal trapezoidal pulse for $g(t)$ with $\Delta_{\rm 1}=\Delta_{\rm 2}$, and $\eta=200$ MHz. Table.~\ref{table:parameters} summarizes the analytical estimates and optimal numerical values for $g_{\rm max}$ and $t_{\rm QST}$.

\begin{table}[htb]
\centering
\caption{Parameters for the control-pulse and the corresponding fidelities (defined in Eq.(\ref{eq:fidelityDef})). Analytical estimates are computed from Eq.(\ref{eq:theoreticalParameters}) and numerical values are obtained via optimization of $g_{\rm max}$ and $t_{\rm QST}$ independently.}
\begin{center}
\resizebox{0.7\linewidth}{!}{
 \begin{tabular}{| c || c | c |}
  \cline{1-3}
Parameters & \multicolumn{2}{c |}{Values} \\
    \cline{2-3}
           & numerical & analytical \\ \hline
    $g_{\rm max}$ (MHz) & 37.7 & 37.5 \\ \hline
    $t_{\rm QST}$ (ns) & 21.95 & 22 \\ \hline\hline
    ${\mathcal F}\;[\%]$ & 99.996 & 99.992 \\ \hline
\end{tabular}}
\label{table:parameters}
\end{center}
\end{table}

Fig.~\ref{fig:stateTransfer} shows the probabilities of population transfer as a function of time for $\ket{10}\rightarrow\ket{01}$ and $\ket{20}\rightarrow\ket{02}$ transitions under the optimal trapezoidal pulse shown in Fig.~\ref{fig:gPulse}. As mentioned earlier, population transfer in the $\{\ket{10},\ket{01}\}$ subspace is faster than that in the $\{\ket{20},\ket{02}\}$ subspace, and in our protocol we set a specific value for $g_{\rm max}$ such that these transfers occur simultaneously coinciding the first peak for the latter with the second peak for the former case. In contrast with the qubit-qubit state-transfer, this unusual matching is, in fact, necessary for our qutrit-qutrit state-transfer, and probably the only choice that satisfies current experimental constraints for superconducting devices. The oscillation observed for the $\ket{20}\rightarrow\ket{02}$ transition in Fig.~\ref{fig:stateTransfer} is due to the interference with the $\ket{11}$ state in the double-excitation subspace.

\subsection{Compensating phases}

In the population transfer protocol described above, the double excitation subspace acquires a phase (in the rotating frame), $\varphi={\eta}t_{\rm QST}$, with respect to the $\{\ket{00},\ket{01},\ket{10}\}$ subspace. Our state-transfer protocol, therefore, consists of the population-transfer plus compensating the additional phases acquired by any of the basis states. Here we discuss how to compensate any arbitrary phase acquired by a superconducting qutrit. The Hamiltonian for a single superconducting qutrit in a rotating frame is given by (in the computational basis),
\BEq
\widetilde{H}_{\rm q}(t)=\left[\begin{array}{ccc}
0 & 0 & 0 \\
0 & \Delta(t) & 0 \\
0 & 0 & 2\Delta(t)-\eta \end{array} \right].
\EEq
In order to perform an arbitrary phase rotation,
\BEq
U_{\rm phase}=\left[\begin{array}{ccc}
1 & 0 & 0 \\
0 & e^{-i\theta} & 0 \\
0 & 0 & e^{-i\phi} \end{array} \right],
\EEq
on the single-qutrit basis states, we vary the time-dependent qutrit-frequency such that,
\BEq
\label{eq:phaseConstraints1}
\begin{array}{l}
\displaystyle \theta=\int\limits_{0}^{t_{\rm phase}}\Delta(t) dt \\
\displaystyle \phi=\int\limits_{0}^{t_{\rm phase}}\left(2\Delta(t)-\eta\right) dt.
\end{array}
\EEq
Eq.(\ref{eq:phaseConstraints1}) is satisfied if we set,
\BEq
\label{eq:phaseConstraints2}
\begin{array}{l}
\displaystyle t_{\rm phase}=\frac{2\theta-\phi}{\eta} \\
\displaystyle \Delta_{\rm max}=\frac{\eta\theta}{2\theta-\phi-2\eta},
\end{array}
\EEq
assuming a trapezoidal pulse for $\Delta(t)$ with $2$ ns ramp, and $\Delta_{\rm max}$ being the maximum value. Eq.(\ref{eq:phaseConstraints2}) can always be satisfied with a proper choice of $\theta$ and $\phi$ modulo $2\pi$.

\subsection{State-transfer fidelity}

The state transfer considered in this section requires one qutrit to be in an arbitrary state $\ket{\psi}$, while the other qutrit is in $\ket{0}$ state. The state transfer operation $U_{\rm QST}$ can, therefore, be represented in matrix form in the basis,
\BEq
\label{eq:basisDef}
\{\ket{00},\ket{01},\ket{10},\ket{02},\ket{20}\}
\EEq
as,
\BEq
U_{\rm QST}=\left[\begin{array}{ccccc}
1 & 0 & 0 & 0 & 0 \\
0 & 0 & 1 & 0 & 0 \\
0 & 1 & 0 & 0 & 0 \\
0 & 0 & 0 & 0 & 1 \\
0 & 0 & 0 & 1 & 0 \end{array} \right].
\EEq
If $U_{\rm obt}$ be the time-evolution operator obtained under the control-pulse shown in Fig.~\ref{fig:gPulse}, then the fidelity ($\mathcal F$) between $U_{\rm obt}$ and $U_{\rm QST}$ is defined as~\cite{PhysRevA.87.022309},
\BEq
\label{eq:fidelityDef}
{\mathcal F}=\frac{{\rm Tr}\left(\widehat{{\mathcal P}}U_{\rm obt}\,U_{\rm obt}^{\dagger}\widehat{{\mathcal P}}\right)+\left|{\rm Tr}\left(U_{\rm QST}^{\dagger}\,\widehat{{\mathcal P}}U_{\rm obt}\widehat{{\mathcal P}}\right)\right|^{2}}{d(d+1)},
\EEq
where $\widehat{{\mathcal P}}$ is the projection operator that projects the time-evolution operator $U_{\rm obt}$ into the computational subspace (\ref{eq:basisDef}), and $d$ is the dimension of the computational subspace, which is $5$ for this case. In absence of decoherence, the dominant source of error in state transfer is the leakage to the $\ket{11}$ state in the double excitation subspace~\cite{PhysRevA.88.062329}, while the phase compensation operation is exact under the model considered for this work. We, therefore, can replace $U_{\rm obt}$ by $\left|U_{\rm obt}\right|$ in Eq.(\ref{eq:fidelityDef}) and compute ${\mathcal F}$ that characterizes the fidelity for both, the state-transfer as well as the population-transfer.

\section{State transfer across a chain of nearest-neighbor-coupled qutrits}
\label{sec:populationTransfer}
Here we describe the model for an array of nearest-neighbor-coupled transmons and then discuss the QST across the chain of transmon quirts.

\subsection{Array of coupled qutrits}
\label{sec:qutritChainModel}
Following the same technique as adopted in Sec.~\ref{sec:coupledQutrit} to derive the coupled-qutrit Hamiltonian (\ref{eq:Htilde}), we can show that the Hamiltonian for a system of $N$ nearest-neighbor-coupled superconducting qutrits is given by (from rotating frame),
\BEqA
\widetilde{H}_{N}(t)&=&\sum\limits_{k=1}^{N} \left[ \begin{array}{ccc}
0 & 0 & 0 \\
0 & \Delta_{\rm k}(t) & 0 \\
0 & 0 & 2\Delta_{\rm k}(t)-\eta \end{array} \right]_{{\rm q}_{k}} \nonumber \\
&+&\sum\limits_{k=1}^{N-1} \frac{g_{k}(t)}{2}\left(X_{k}X_{k+1}+Y_{k}Y_{k+1}\right),
\label{eq:QCMHamiltonian}
\EEqA
where $\Delta_{k}$ is frequency of $k^{\rm th}$ transmon measured in reference to the frequency of the rotating frame, and $X_{k}$ and $Y_{k}$ are three-dimensional generalizations of Pauli's $\sigma^{x}$ and $\sigma^{y}$ matrices (corresponding to the $k^{\rm th}$ qutrit), as defined in Eq.(\ref{eq:defX}) and Eq.(\ref{eq:defY}) respectively. While both the frequencies and coupling strengths are time-dependent for our system, in order to perform QST we keep all the qutrits in resonance, i.e., $\Delta_{k}=0,\,\; \forall k \in \{1,2, \ldots, N\}$, and control the coupling strengths $g_{k}$ with external control pulses.

Our QST protocol is composed of sequential state-transfer steps between adjacent qutrits, which means for $N$ qutrits we need to perform $N-1$ sequential QST operations. It is, therefore, equivalent if we explore the accumulation of error for our protocol as a function of number of qutrits or as a function of number of concatenated state-transfer steps. We here adopt the latter and analyze the error mechanisms for our approach in the next section.

\subsection{State transfer protocol}
\begin{figure}[htb]
\centering
\includegraphics[angle=0,width=0.95\linewidth]{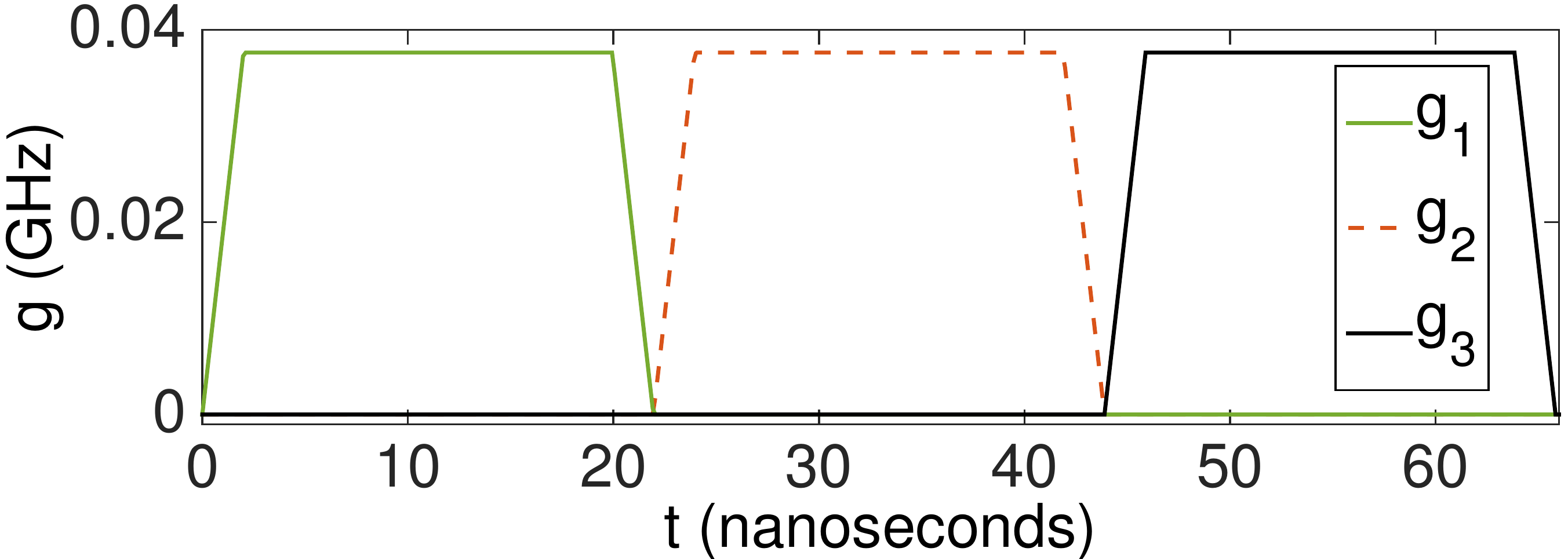}
\caption{(Color online) Trapezoidal pulses for $g_{k}(t)$ for a state-transfer across a chain of $4$ nearest-neighbor-coupled superconducting qutrits. In order to emulate a QST across a chain of $N$ qutrits, we need to concatenate $(N-1)$ such pulses.}
\label{fig:gVst}
\end{figure}
As mentioned earlier, all the qutrits are always in resonance during our QST protocol, while the coupling strengths are changed sequentially to transfer our initial state successively from one qutrit to another via neighboring qutrits. Fig.~\ref{fig:gVst} shows our sequential trapezoidal control pulses for a QST across a chain of $4$ coupled qutrits, where we use the optimal parameters (shown in Table.~\ref{table:parameters}) obtained numerically for the two-qutrit state transfer. A state-transfer across a chain of $N$ coupled qutrits requires concatenation of $(N-1)$ such pulses one after another, as mentioned earlier. We emphasize that, it is sufficient for our QST protocol if we just optimize the pulse for a single qutrit-qutrit state transfer, and then combine the pulses sequentially as shown in Fig.~\ref{fig:gVst}. This modularity is, in fact, required for any scalable QST protocol. 

\section{Analysis of errors}
\label{sec:errors}
Here we discuss various error-mechanisms relevant for our QST scheme. First, we estimate the errors generated from the unitary evolution under the control pulse (intrinsic errors), and then explore the effect of decoherence.

\subsection{Intrinsic errors}

\begin{figure}[htb]
\centering
\includegraphics[angle=0,width=0.95\linewidth]{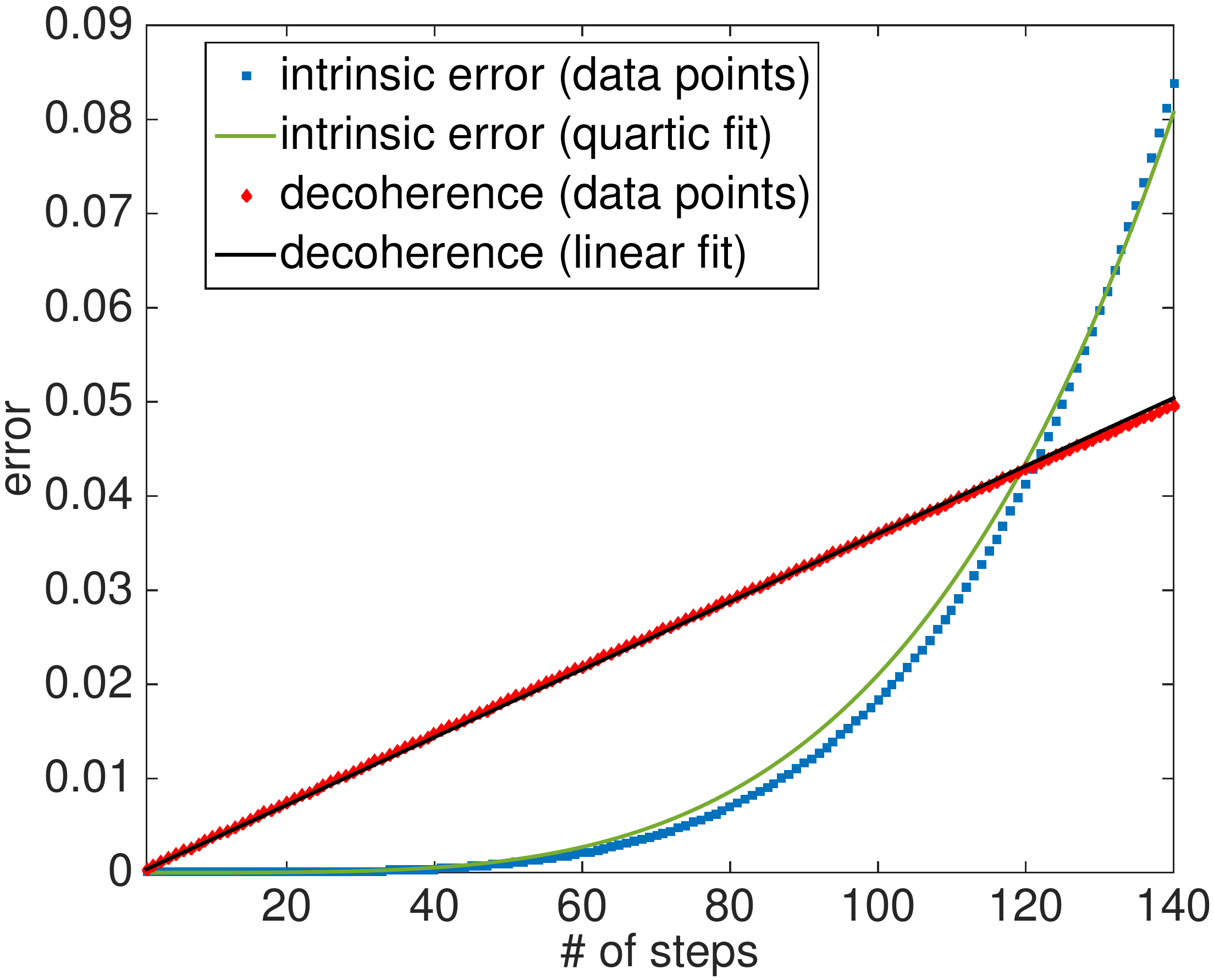}
\caption{(Color online) Accumulation of intrinsic and decoherence-induced errors with the number of steps. The red diamonds and blue squares are numerically computed data-points, and the solid black and green (gray) curves are the linear and quartic fit for the decoherence-induced and intrinsic errors respectively.}
\label{fig:errors}
\end{figure}

Our QST scheme is composed of concatenating successive trapezoidal pulses for the coupling strengths, where the same set of optimal parameters is used for each pulse. Intrinsic errors are defined as errors originating from the unitary evolution of the system under the control pulse at $T_{\rm 1,2}\rightarrow\infty$ limit. In order to quantify how the intrinsic errors accumulate with sequential state-transfer steps, we prepare a uniform superposition $\psi_{\rm unif}=(\ket{0}+\ket{1}+\ket{2})/{\sqrt{3}}$ in the first qutrit, and then compute the error after every state-transfer step to the adjacent qutrit. If $\psi_k$ is the quantum state transferred at the $k^{\rm th}$ step to the $(k+1)^{\rm th}$ qutrit, then we define the intrinsic error as,
\BEq
{\mathcal E}^{\rm intr}_{k}=1-\left|\braket{\psi_{\rm unif}}{\psi_k}\right|^{2}.
\EEq
The blue (square) data-points in Fig.~\ref{fig:errors} show the intrinsic error as a function of the number of steps, and we observe a quartic accumulation of intrinsic errors in that regime. The green (gray) curve in Fig.~\ref{fig:errors} is a quartic fit corresponding to ${\mathcal E}^{\rm intr}_{k}={\mathcal A}k^4$, where the pre-factor ${\mathcal A}$ is numerically determined to be $\sim2.1\times{10^{-10}}$ for our case. The quartic accumulation of intrinsic errors, as opposed to an exponential accumulation~\cite{PhysRevA.83.012325}, in fact allows us to perform a state-transfer across a longer chain of superconducting qutrits.

It should be emphasized at this point, that many other error mechanisms can occur in a realistic setup, such as errors generated by the imperfect control electronics. Also, one can design different pulse shapes satisfying the constraint derived in this work, and imperfection in concatenating various pulse shapes can generate considerable intrinsic errors. While the robustness of our approach against such realistic noise-mechanisms could be a topic of future research, we here consider a perfect experimental control and concatenation, and analyze the intrinsic error that comes from the leakage of population into some undesired states. 

\subsection{Effects of decoherence}

The model considered for this work assumes tunable couplings between adjacent qutrits, which means during the entire state-transfer all the qutrits are decoupled from the system as well as remain in the ground state, except for the two neighboring qutrits participating in the QST. We, therefore, argue that the effects of decoherence on the qutrit state is essentially equivalent to that on a single qutrit prepared in the same state during the entire state-transfer process. In order to quantify the decoherence-induced errors on our QST scheme, we consider a single qutrit prepared in a uniform superposition $\psi_{\rm unif}=(\ket{0}+\ket{1}+\ket{2})/{\sqrt{3}}$ (as considered for estimating the intrinsic errors), and construct the Kraus matrices for the amplitude and phase damping using the damped harmonic oscillator approximation~\cite{PhysRevA.70.042308}. We then perform the Kraus evolution for a time-duration $kt_{\rm QST}$ (time required for $k$ successive state-transfer steps) on the single-qutrit density matrix $\rho$. The red (diamond-shaped) data-points in Fig.~\ref{fig:errors} show the decoherence-induced error,
\BEq
{\mathcal E}^{\rm decoh}_{k}=1-\bra{\psi_{\rm unif}}\rho(kt_{\rm QST})\ket{\psi_{\rm unif}},
\EEq
as a function of $k$. The black line (almost aligned with the blue data-points) in Fig.~\ref{fig:errors} shows the linear fit for the decoherence-induced error corresponding to ${\mathcal E}^{\rm decoh}_{k}={\mathcal B}k$, where the pre-factor ${\mathcal B}$ is numerically determined to be $\sim3.6\times10^{-4}$. This numerical estimate of the slope of the linear fit in Fig.~\ref{fig:errors} is consistent with the approximate analytical estimate $t_{\rm QST}/T_{\rm 1,2}$ ($\approx3.66\times10^{-4}$), where we assume $T_{\rm 1}=T_{\rm 2}=60$ $\mu$s for the superconducting transmon qutrits \cite{Barends2014}. It is interesting to note that for our case decoherence is dominated by the intrinsic errors for $k>120$, due the the quartic scaling of the intrinsic errors.

\section{Conclusions}
\label{sec:conclusions}
In this work, we have introduced a proposal for emulating a QST across a chain of spin-1 systems on a lattice of nearest-neighbor-coupled superconducting qutrits. While the emulation of higher spin systems with a single superconducting artificial atom has been demonstrated earlier~\cite{Neeley07082009}, the problem transmitting a qutrit state along a chain of superconducting atoms has remained a nontrivial problem primarily due to the unequal coupling strengths in the single- and double-excitation subspaces. Here we have shown how to overcome this challenge with a proper choice of the control parameters under existing experimental conditions. Our proposal thus motivates the simulation of various quantum transport processes across higher spin systems, as well as enhanced quantum communication with scalable superconducting qutrits. Some possible future directions of this work include transmission of an arbitrary \emph{qudit} state (a state encoded in a $d$-level quantum system) along a chain of coupled superconducting atoms and transfer of various entangled qutrit states across a chain of superconducting qutrits.

\begin{acknowledgments}
This research was funded by NSERC, AITF and University of Calgary's Eyes High Fellowship Program. I thank Barry Sanders for many illuminating comments as well as his careful reading of the manuscript. I also gratefully acknowledge useful discussions with David Feder, Michael Geller and Pedram Roushan.
\end{acknowledgments}

\bibliography{qutritStateTransfer}

\begin{thebibliography}{36}%
\makeatletter
\providecommand \@ifxundefined [1]{%
 \@ifx{#1\undefined}
}%
\providecommand \@ifnum [1]{%
 \ifnum #1\expandafter \@firstoftwo
 \else \expandafter \@secondoftwo
 \fi
}%
\providecommand \@ifx [1]{%
 \ifx #1\expandafter \@firstoftwo
 \else \expandafter \@secondoftwo
 \fi
}%
\providecommand \natexlab [1]{#1}%
\providecommand \enquote  [1]{``#1''}%
\providecommand \bibnamefont  [1]{#1}%
\providecommand \bibfnamefont [1]{#1}%
\providecommand \citenamefont [1]{#1}%
\providecommand \href@noop [0]{\@secondoftwo}%
\providecommand \href [0]{\begingroup \@sanitize@url \@href}%
\providecommand \@href[1]{\@@startlink{#1}\@@href}%
\providecommand \@@href[1]{\endgroup#1\@@endlink}%
\providecommand \@sanitize@url [0]{\catcode `\\12\catcode `\$12\catcode
  `\&12\catcode `\#12\catcode `\^12\catcode `\_12\catcode `\%12\relax}%
\providecommand \@@startlink[1]{}%
\providecommand \@@endlink[0]{}%
\providecommand \url  [0]{\begingroup\@sanitize@url \@url }%
\providecommand \@url [1]{\endgroup\@href {#1}{\urlprefix }}%
\providecommand \urlprefix  [0]{URL }%
\providecommand \Eprint [0]{\href }%
\providecommand \doibase [0]{http://dx.doi.org/}%
\providecommand \selectlanguage [0]{\@gobble}%
\providecommand \bibinfo  [0]{\@secondoftwo}%
\providecommand \bibfield  [0]{\@secondoftwo}%
\providecommand \translation [1]{[#1]}%
\providecommand \BibitemOpen [0]{}%
\providecommand \bibitemStop [0]{}%
\providecommand \bibitemNoStop [0]{.\EOS\space}%
\providecommand \EOS [0]{\spacefactor3000\relax}%
\providecommand \BibitemShut  [1]{\csname bibitem#1\endcsname}%
\let\auto@bib@innerbib\@empty
\bibitem [{\citenamefont {Bose}(2003)}]{PhysRevLett.91.207901}%
  \BibitemOpen
  \bibfield  {author} {\bibinfo {author} {\bibfnamefont {S.}~\bibnamefont
  {Bose}},\ }\href {\doibase 10.1103/PhysRevLett.91.207901} {\bibfield
  {journal} {\bibinfo  {journal} {Phys. Rev. Lett.}\ }\textbf {\bibinfo
  {volume} {91}},\ \bibinfo {pages} {207901} (\bibinfo {year}
  {2003})}\BibitemShut {NoStop}%
\bibitem [{\citenamefont {Albanese}\ \emph {et~al.}(2004)\citenamefont
  {Albanese}, \citenamefont {Christandl}, \citenamefont {Datta},\ and\
  \citenamefont {Ekert}}]{PhysRevLett.93.230502}%
  \BibitemOpen
  \bibfield  {author} {\bibinfo {author} {\bibfnamefont {C.}~\bibnamefont
  {Albanese}}, \bibinfo {author} {\bibfnamefont {M.}~\bibnamefont
  {Christandl}}, \bibinfo {author} {\bibfnamefont {N.}~\bibnamefont {Datta}}, \
  and\ \bibinfo {author} {\bibfnamefont {A.}~\bibnamefont {Ekert}},\ }\href
  {\doibase 10.1103/PhysRevLett.93.230502} {\bibfield  {journal} {\bibinfo
  {journal} {Phys. Rev. Lett.}\ }\textbf {\bibinfo {volume} {93}},\ \bibinfo
  {pages} {230502} (\bibinfo {year} {2004})}\BibitemShut {NoStop}%
\bibitem [{\citenamefont {Subrahmanyam}(2004)}]{PhysRevA.69.034304}%
  \BibitemOpen
  \bibfield  {author} {\bibinfo {author} {\bibfnamefont {V.}~\bibnamefont
  {Subrahmanyam}},\ }\href {\doibase 10.1103/PhysRevA.69.034304} {\bibfield
  {journal} {\bibinfo  {journal} {Phys. Rev. A}\ }\textbf {\bibinfo {volume}
  {69}},\ \bibinfo {pages} {034304} (\bibinfo {year} {2004})}\BibitemShut
  {NoStop}%
\bibitem [{\citenamefont {Korzekwa}\ \emph {et~al.}(2014)\citenamefont
  {Korzekwa}, \citenamefont {Machnikowski},\ and\ \citenamefont
  {Horodecki}}]{PhysRevA.89.062301}%
  \BibitemOpen
  \bibfield  {author} {\bibinfo {author} {\bibfnamefont {K.}~\bibnamefont
  {Korzekwa}}, \bibinfo {author} {\bibfnamefont {P.}~\bibnamefont
  {Machnikowski}}, \ and\ \bibinfo {author} {\bibfnamefont {P.}~\bibnamefont
  {Horodecki}},\ }\href {\doibase 10.1103/PhysRevA.89.062301} {\bibfield
  {journal} {\bibinfo  {journal} {Phys. Rev. A}\ }\textbf {\bibinfo {volume}
  {89}},\ \bibinfo {pages} {062301} (\bibinfo {year} {2014})}\BibitemShut
  {NoStop}%
\bibitem [{\citenamefont {Di~Franco}\ \emph {et~al.}(2008)\citenamefont
  {Di~Franco}, \citenamefont {Paternostro},\ and\ \citenamefont
  {Kim}}]{PhysRevLett.101.230502}%
  \BibitemOpen
  \bibfield  {author} {\bibinfo {author} {\bibfnamefont {C.}~\bibnamefont
  {Di~Franco}}, \bibinfo {author} {\bibfnamefont {M.}~\bibnamefont
  {Paternostro}}, \ and\ \bibinfo {author} {\bibfnamefont {M.~S.}\ \bibnamefont
  {Kim}},\ }\href {\doibase 10.1103/PhysRevLett.101.230502} {\bibfield
  {journal} {\bibinfo  {journal} {Phys. Rev. Lett.}\ }\textbf {\bibinfo
  {volume} {101}},\ \bibinfo {pages} {230502} (\bibinfo {year}
  {2008})}\BibitemShut {NoStop}%
\bibitem [{\citenamefont {Bose}(2007)}]{doi:10.1080/00107510701342313}%
  \BibitemOpen
  \bibfield  {author} {\bibinfo {author} {\bibfnamefont {S.}~\bibnamefont
  {Bose}},\ }\href {\doibase 10.1080/00107510701342313} {\bibfield  {journal}
  {\bibinfo  {journal} {Contemporary Physics}\ }\textbf {\bibinfo {volume}
  {48}},\ \bibinfo {pages} {13} (\bibinfo {year} {2007})}\BibitemShut {NoStop}%
\bibitem [{\citenamefont {Kay}(2010)}]{doi:10.1142/S0219749910006514}%
  \BibitemOpen
  \bibfield  {author} {\bibinfo {author} {\bibfnamefont {A.}~\bibnamefont
  {Kay}},\ }\href {\doibase 10.1142/S0219749910006514} {\bibfield  {journal}
  {\bibinfo  {journal} {International Journal of Quantum Information}\ }\textbf
  {\bibinfo {volume} {08}},\ \bibinfo {pages} {641} (\bibinfo {year}
  {2010})}\BibitemShut {NoStop}%
\bibitem [{\citenamefont {Cappellaro}\ \emph {et~al.}(2007)\citenamefont
  {Cappellaro}, \citenamefont {Ramanathan},\ and\ \citenamefont
  {Cory}}]{PhysRevLett.99.250506}%
  \BibitemOpen
  \bibfield  {author} {\bibinfo {author} {\bibfnamefont {P.}~\bibnamefont
  {Cappellaro}}, \bibinfo {author} {\bibfnamefont {C.}~\bibnamefont
  {Ramanathan}}, \ and\ \bibinfo {author} {\bibfnamefont {D.~G.}\ \bibnamefont
  {Cory}},\ }\href {\doibase 10.1103/PhysRevLett.99.250506} {\bibfield
  {journal} {\bibinfo  {journal} {Phys. Rev. Lett.}\ }\textbf {\bibinfo
  {volume} {99}},\ \bibinfo {pages} {250506} (\bibinfo {year}
  {2007})}\BibitemShut {NoStop}%
\bibitem [{\citenamefont {Ajoy}\ \emph {et~al.}(2012)\citenamefont {Ajoy},
  \citenamefont {Rao}, \citenamefont {Kumar},\ and\ \citenamefont
  {Rungta}}]{PhysRevA.85.030303}%
  \BibitemOpen
  \bibfield  {author} {\bibinfo {author} {\bibfnamefont {A.}~\bibnamefont
  {Ajoy}}, \bibinfo {author} {\bibfnamefont {R.~K.}\ \bibnamefont {Rao}},
  \bibinfo {author} {\bibfnamefont {A.}~\bibnamefont {Kumar}}, \ and\ \bibinfo
  {author} {\bibfnamefont {P.}~\bibnamefont {Rungta}},\ }\href {\doibase
  10.1103/PhysRevA.85.030303} {\bibfield  {journal} {\bibinfo  {journal} {Phys.
  Rev. A}\ }\textbf {\bibinfo {volume} {85}},\ \bibinfo {pages} {030303}
  (\bibinfo {year} {2012})}\BibitemShut {NoStop}%
\bibitem [{\citenamefont {Kaur}\ and\ \citenamefont
  {Cappellaro}(2012)}]{1367-2630-14-8-083005}%
  \BibitemOpen
  \bibfield  {author} {\bibinfo {author} {\bibfnamefont {G.}~\bibnamefont
  {Kaur}}\ and\ \bibinfo {author} {\bibfnamefont {P.}~\bibnamefont
  {Cappellaro}},\ }\href {http://stacks.iop.org/1367-2630/14/i=8/a=083005}
  {\bibfield  {journal} {\bibinfo  {journal} {New Journal of Physics}\ }\textbf
  {\bibinfo {volume} {14}},\ \bibinfo {pages} {083005} (\bibinfo {year}
  {2012})}\BibitemShut {NoStop}%
\bibitem [{\citenamefont {M\"{u}ller}\ \emph {et~al.}(2008)\citenamefont
  {M\"{u}ller}, \citenamefont {Liang}, \citenamefont {Lesanovsky},\ and\
  \citenamefont {Zoller}}]{1367-2630-10-9-093009}%
  \BibitemOpen
  \bibfield  {author} {\bibinfo {author} {\bibfnamefont {M.}~\bibnamefont
  {M\"{u}ller}}, \bibinfo {author} {\bibfnamefont {L.}~\bibnamefont {Liang}},
  \bibinfo {author} {\bibfnamefont {I.}~\bibnamefont {Lesanovsky}}, \ and\
  \bibinfo {author} {\bibfnamefont {P.}~\bibnamefont {Zoller}},\ }\href
  {http://stacks.iop.org/1367-2630/10/i=9/a=093009} {\bibfield  {journal}
  {\bibinfo  {journal} {New Journal of Physics}\ }\textbf {\bibinfo {volume}
  {10}},\ \bibinfo {pages} {093009} (\bibinfo {year} {2008})}\BibitemShut
  {NoStop}%
\bibitem [{\citenamefont {Liu}\ and\ \citenamefont
  {Zhou}(2014)}]{liu2014transfer}%
  \BibitemOpen
  \bibfield  {author} {\bibinfo {author} {\bibfnamefont {Y.}~\bibnamefont
  {Liu}}\ and\ \bibinfo {author} {\bibfnamefont {D.}~\bibnamefont {Zhou}},\
  }\href@noop {} {\bibfield  {journal} {\bibinfo  {journal} {arXiv preprint
  arXiv:1405.2634}\ } (\bibinfo {year} {2014})}\BibitemShut {NoStop}%
\bibitem [{\citenamefont {Lyakhov}\ and\ \citenamefont
  {Bruder}(2005)}]{1367-2630-7-1-181}%
  \BibitemOpen
  \bibfield  {author} {\bibinfo {author} {\bibfnamefont {A.}~\bibnamefont
  {Lyakhov}}\ and\ \bibinfo {author} {\bibfnamefont {C.}~\bibnamefont
  {Bruder}},\ }\href {http://stacks.iop.org/1367-2630/7/i=1/a=181} {\bibfield
  {journal} {\bibinfo  {journal} {New Journal of Physics}\ }\textbf {\bibinfo
  {volume} {7}},\ \bibinfo {pages} {181} (\bibinfo {year} {2005})}\BibitemShut
  {NoStop}%
\bibitem [{\citenamefont {Rao}\ \emph {et~al.}(2014)\citenamefont {Rao},
  \citenamefont {Mahesh},\ and\ \citenamefont {Kumar}}]{rao2013simulation}%
  \BibitemOpen
  \bibfield  {author} {\bibinfo {author} {\bibfnamefont {K.~R.~K.}\
  \bibnamefont {Rao}}, \bibinfo {author} {\bibfnamefont {T.~S.}\ \bibnamefont
  {Mahesh}}, \ and\ \bibinfo {author} {\bibfnamefont {A.}~\bibnamefont
  {Kumar}},\ }\href {\doibase 10.1103/PhysRevA.90.012306} {\bibfield  {journal}
  {\bibinfo  {journal} {Phys. Rev. A}\ }\textbf {\bibinfo {volume} {90}},\
  \bibinfo {pages} {012306} (\bibinfo {year} {2014})}\BibitemShut {NoStop}%
\bibitem [{\citenamefont {Bellec}\ \emph {et~al.}(2012)\citenamefont {Bellec},
  \citenamefont {Nikolopoulos},\ and\ \citenamefont {Tzortzakis}}]{Bellec:12}%
  \BibitemOpen
  \bibfield  {author} {\bibinfo {author} {\bibfnamefont {M.}~\bibnamefont
  {Bellec}}, \bibinfo {author} {\bibfnamefont {G.~M.}\ \bibnamefont
  {Nikolopoulos}}, \ and\ \bibinfo {author} {\bibfnamefont {S.}~\bibnamefont
  {Tzortzakis}},\ }\href {\doibase 10.1364/OL.37.004504} {\bibfield  {journal}
  {\bibinfo  {journal} {Opt. Lett.}\ }\textbf {\bibinfo {volume} {37}},\
  \bibinfo {pages} {4504} (\bibinfo {year} {2012})}\BibitemShut {NoStop}%
\bibitem [{\citenamefont {Perez-Leija}\ \emph {et~al.}(2013)\citenamefont
  {Perez-Leija}, \citenamefont {Keil}, \citenamefont {Kay}, \citenamefont
  {Moya-Cessa}, \citenamefont {Nolte}, \citenamefont {Kwek}, \citenamefont
  {Rodr\'{i}guez-Lara}, \citenamefont {Szameit},\ and\ \citenamefont
  {Christodoulides}}]{PhysRevA.87.012309}%
  \BibitemOpen
  \bibfield  {author} {\bibinfo {author} {\bibfnamefont {A.}~\bibnamefont
  {Perez-Leija}}, \bibinfo {author} {\bibfnamefont {R.}~\bibnamefont {Keil}},
  \bibinfo {author} {\bibfnamefont {A.}~\bibnamefont {Kay}}, \bibinfo {author}
  {\bibfnamefont {H.}~\bibnamefont {Moya-Cessa}}, \bibinfo {author}
  {\bibfnamefont {S.}~\bibnamefont {Nolte}}, \bibinfo {author} {\bibfnamefont
  {L.-C.}\ \bibnamefont {Kwek}}, \bibinfo {author} {\bibfnamefont {B.~M.}\
  \bibnamefont {Rodr\'{i}guez-Lara}}, \bibinfo {author} {\bibfnamefont
  {A.}~\bibnamefont {Szameit}}, \ and\ \bibinfo {author} {\bibfnamefont
  {D.~N.}\ \bibnamefont {Christodoulides}},\ }\href {\doibase
  10.1103/PhysRevA.87.012309} {\bibfield  {journal} {\bibinfo  {journal} {Phys.
  Rev. A}\ }\textbf {\bibinfo {volume} {87}},\ \bibinfo {pages} {012309}
  (\bibinfo {year} {2013})}\BibitemShut {NoStop}%
\bibitem [{\citenamefont {Fukuhara}\ \emph
  {et~al.}(2013{\natexlab{a}})\citenamefont {Fukuhara}, \citenamefont
  {Kantian}, \citenamefont {Endres}, \citenamefont {Cheneau}, \citenamefont
  {Schausz}, \citenamefont {Hild}, \citenamefont {Bellem}, \citenamefont
  {Schollwock}, \citenamefont {Giamarchi}, \citenamefont {Gross}, \citenamefont
  {Bloch},\ and\ \citenamefont {Kuhr}}]{Fukuhara2013a}%
  \BibitemOpen
  \bibfield  {author} {\bibinfo {author} {\bibfnamefont {T.}~\bibnamefont
  {Fukuhara}}, \bibinfo {author} {\bibfnamefont {A.}~\bibnamefont {Kantian}},
  \bibinfo {author} {\bibfnamefont {M.}~\bibnamefont {Endres}}, \bibinfo
  {author} {\bibfnamefont {M.}~\bibnamefont {Cheneau}}, \bibinfo {author}
  {\bibfnamefont {P.}~\bibnamefont {Schausz}}, \bibinfo {author} {\bibfnamefont
  {S.}~\bibnamefont {Hild}}, \bibinfo {author} {\bibfnamefont {D.}~\bibnamefont
  {Bellem}}, \bibinfo {author} {\bibfnamefont {U.}~\bibnamefont {Schollwock}},
  \bibinfo {author} {\bibfnamefont {T.}~\bibnamefont {Giamarchi}}, \bibinfo
  {author} {\bibfnamefont {C.}~\bibnamefont {Gross}}, \bibinfo {author}
  {\bibfnamefont {I.}~\bibnamefont {Bloch}}, \ and\ \bibinfo {author}
  {\bibfnamefont {S.}~\bibnamefont {Kuhr}},\ }\href
  {http://dx.doi.org/10.1038/nphys2561} {\bibfield  {journal} {\bibinfo
  {journal} {Nat Phys}\ }\textbf {\bibinfo {volume} {9}},\ \bibinfo {pages}
  {235} (\bibinfo {year} {2013}{\natexlab{a}})}\BibitemShut {NoStop}%
\bibitem [{\citenamefont {Fukuhara}\ \emph
  {et~al.}(2013{\natexlab{b}})\citenamefont {Fukuhara}, \citenamefont
  {Schausz}, \citenamefont {Endres}, \citenamefont {Hild}, \citenamefont
  {Cheneau}, \citenamefont {Bloch},\ and\ \citenamefont
  {Gross}}]{Fukuhara2013}%
  \BibitemOpen
  \bibfield  {author} {\bibinfo {author} {\bibfnamefont {T.}~\bibnamefont
  {Fukuhara}}, \bibinfo {author} {\bibfnamefont {P.}~\bibnamefont {Schausz}},
  \bibinfo {author} {\bibfnamefont {M.}~\bibnamefont {Endres}}, \bibinfo
  {author} {\bibfnamefont {S.}~\bibnamefont {Hild}}, \bibinfo {author}
  {\bibfnamefont {M.}~\bibnamefont {Cheneau}}, \bibinfo {author} {\bibfnamefont
  {I.}~\bibnamefont {Bloch}}, \ and\ \bibinfo {author} {\bibfnamefont
  {C.}~\bibnamefont {Gross}},\ }\href {http://dx.doi.org/10.1038/nature12541}
  {\bibfield  {journal} {\bibinfo  {journal} {Nature}\ }\textbf {\bibinfo
  {volume} {502}},\ \bibinfo {pages} {76} (\bibinfo {year}
  {2013}{\natexlab{b}})}\BibitemShut {NoStop}%
\bibitem [{\citenamefont {Bechmann-Pasquinucci}\ and\ \citenamefont
  {Tittel}(2000)}]{PhysRevA.61.062308}%
  \BibitemOpen
  \bibfield  {author} {\bibinfo {author} {\bibfnamefont {H.}~\bibnamefont
  {Bechmann-Pasquinucci}}\ and\ \bibinfo {author} {\bibfnamefont
  {W.}~\bibnamefont {Tittel}},\ }\href {\doibase 10.1103/PhysRevA.61.062308}
  {\bibfield  {journal} {\bibinfo  {journal} {Phys. Rev. A}\ }\textbf {\bibinfo
  {volume} {61}},\ \bibinfo {pages} {062308} (\bibinfo {year}
  {2000})}\BibitemShut {NoStop}%
\bibitem [{\citenamefont {Durt}\ \emph {et~al.}(2003)\citenamefont {Durt},
  \citenamefont {Cerf}, \citenamefont {Gisin},\ and\ \citenamefont
  {\ifmmode~\dot{Z}\else \.{Z}\fi{}ukowski}}]{PhysRevA.67.012311}%
  \BibitemOpen
  \bibfield  {author} {\bibinfo {author} {\bibfnamefont {T.}~\bibnamefont
  {Durt}}, \bibinfo {author} {\bibfnamefont {N.~J.}\ \bibnamefont {Cerf}},
  \bibinfo {author} {\bibfnamefont {N.}~\bibnamefont {Gisin}}, \ and\ \bibinfo
  {author} {\bibfnamefont {M.}~\bibnamefont {\ifmmode~\dot{Z}\else
  \.{Z}\fi{}ukowski}},\ }\href {\doibase 10.1103/PhysRevA.67.012311} {\bibfield
   {journal} {\bibinfo  {journal} {Phys. Rev. A}\ }\textbf {\bibinfo {volume}
  {67}},\ \bibinfo {pages} {012311} (\bibinfo {year} {2003})}\BibitemShut
  {NoStop}%
\bibitem [{\citenamefont {Eckert}\ \emph {et~al.}(2007)\citenamefont {Eckert},
  \citenamefont {Romero-Isart},\ and\ \citenamefont
  {Sanpera}}]{1367-2630-9-5-155}%
  \BibitemOpen
  \bibfield  {author} {\bibinfo {author} {\bibfnamefont {K.}~\bibnamefont
  {Eckert}}, \bibinfo {author} {\bibfnamefont {O.}~\bibnamefont
  {Romero-Isart}}, \ and\ \bibinfo {author} {\bibfnamefont {A.}~\bibnamefont
  {Sanpera}},\ }\href {http://stacks.iop.org/1367-2630/9/i=5/a=155} {\bibfield
  {journal} {\bibinfo  {journal} {New Journal of Physics}\ }\textbf {\bibinfo
  {volume} {9}},\ \bibinfo {pages} {155} (\bibinfo {year} {2007})}\BibitemShut
  {NoStop}%
\bibitem [{\citenamefont {Qin}\ \emph {et~al.}(2013)\citenamefont {Qin},
  \citenamefont {Wang},\ and\ \citenamefont {Long}}]{PhysRevA.87.012339}%
  \BibitemOpen
  \bibfield  {author} {\bibinfo {author} {\bibfnamefont {W.}~\bibnamefont
  {Qin}}, \bibinfo {author} {\bibfnamefont {C.}~\bibnamefont {Wang}}, \ and\
  \bibinfo {author} {\bibfnamefont {G.~L.}\ \bibnamefont {Long}},\ }\href
  {\doibase 10.1103/PhysRevA.87.012339} {\bibfield  {journal} {\bibinfo
  {journal} {Phys. Rev. A}\ }\textbf {\bibinfo {volume} {87}},\ \bibinfo
  {pages} {012339} (\bibinfo {year} {2013})}\BibitemShut {NoStop}%
\bibitem [{\citenamefont {Asoudeh}\ and\ \citenamefont
  {Karimipour}(2014)}]{Asoudeh2014QIPraey}%
  \BibitemOpen
  \bibfield  {author} {\bibinfo {author} {\bibfnamefont {M.}~\bibnamefont
  {Asoudeh}}\ and\ \bibinfo {author} {\bibfnamefont {V.}~\bibnamefont
  {Karimipour}},\ }\href {\doibase 10.1007/s11128-013-0676-8} {\bibfield
  {journal} {\bibinfo  {journal} {Quantum Information Processing}\ }\textbf
  {\bibinfo {volume} {13}},\ \bibinfo {pages} {601} (\bibinfo {year}
  {2014})}\BibitemShut {NoStop}%
\bibitem [{\citenamefont {Bayat}(2014)}]{PhysRevA.89.062302}%
  \BibitemOpen
  \bibfield  {author} {\bibinfo {author} {\bibfnamefont {A.}~\bibnamefont
  {Bayat}},\ }\href {\doibase 10.1103/PhysRevA.89.062302} {\bibfield  {journal}
  {\bibinfo  {journal} {Phys. Rev. A}\ }\textbf {\bibinfo {volume} {89}},\
  \bibinfo {pages} {062302} (\bibinfo {year} {2014})}\BibitemShut {NoStop}%
\bibitem [{\citenamefont {Romero-Isart}\ \emph {et~al.}(2007)\citenamefont
  {Romero-Isart}, \citenamefont {Eckert},\ and\ \citenamefont
  {Sanpera}}]{PhysRevA.75.050303}%
  \BibitemOpen
  \bibfield  {author} {\bibinfo {author} {\bibfnamefont {O.}~\bibnamefont
  {Romero-Isart}}, \bibinfo {author} {\bibfnamefont {K.}~\bibnamefont
  {Eckert}}, \ and\ \bibinfo {author} {\bibfnamefont {A.}~\bibnamefont
  {Sanpera}},\ }\href {\doibase 10.1103/PhysRevA.75.050303} {\bibfield
  {journal} {\bibinfo  {journal} {Phys. Rev. A}\ }\textbf {\bibinfo {volume}
  {75}},\ \bibinfo {pages} {050303} (\bibinfo {year} {2007})}\BibitemShut
  {NoStop}%
\bibitem [{\citenamefont {Wiesniak}\ \emph {et~al.}(2013)\citenamefont
  {Wiesniak}, \citenamefont {Dutta},\ and\ \citenamefont
  {Ryu}}]{wiesniak2013translating}%
  \BibitemOpen
  \bibfield  {author} {\bibinfo {author} {\bibfnamefont {M.}~\bibnamefont
  {Wiesniak}}, \bibinfo {author} {\bibfnamefont {A.}~\bibnamefont {Dutta}}, \
  and\ \bibinfo {author} {\bibfnamefont {J.}~\bibnamefont {Ryu}},\ }\href@noop
  {} {\bibfield  {journal} {\bibinfo  {journal} {arXiv preprint
  arXiv:1312.6543}\ } (\bibinfo {year} {2013})}\BibitemShut {NoStop}%
\bibitem [{\citenamefont {Delgado}\ \emph {et~al.}(2007)\citenamefont
  {Delgado}, \citenamefont {Saavedra},\ and\ \citenamefont
  {Retamal}}]{Delgado200722}%
  \BibitemOpen
  \bibfield  {author} {\bibinfo {author} {\bibfnamefont {A.}~\bibnamefont
  {Delgado}}, \bibinfo {author} {\bibfnamefont {C.}~\bibnamefont {Saavedra}}, \
  and\ \bibinfo {author} {\bibfnamefont {J.}~\bibnamefont {Retamal}},\ }\href
  {\doibase http://dx.doi.org/10.1016/j.physleta.2007.05.022} {\bibfield
  {journal} {\bibinfo  {journal} {Physics Letters A}\ }\textbf {\bibinfo
  {volume} {370}},\ \bibinfo {pages} {22 } (\bibinfo {year}
  {2007})}\BibitemShut {NoStop}%
\bibitem [{\citenamefont {Neeley}\ \emph {et~al.}(2009)\citenamefont {Neeley},
  \citenamefont {Ansmann}, \citenamefont {Bialczak}, \citenamefont {Hofheinz},
  \citenamefont {Lucero}, \citenamefont {O'Connell}, \citenamefont {Sank},
  \citenamefont {Wang}, \citenamefont {Wenner}, \citenamefont {Cleland},
  \citenamefont {Geller},\ and\ \citenamefont {Martinis}}]{Neeley07082009}%
  \BibitemOpen
  \bibfield  {author} {\bibinfo {author} {\bibfnamefont {M.}~\bibnamefont
  {Neeley}}, \bibinfo {author} {\bibfnamefont {M.}~\bibnamefont {Ansmann}},
  \bibinfo {author} {\bibfnamefont {R.~C.}\ \bibnamefont {Bialczak}}, \bibinfo
  {author} {\bibfnamefont {M.}~\bibnamefont {Hofheinz}}, \bibinfo {author}
  {\bibfnamefont {E.}~\bibnamefont {Lucero}}, \bibinfo {author} {\bibfnamefont
  {A.~D.}\ \bibnamefont {O'Connell}}, \bibinfo {author} {\bibfnamefont
  {D.}~\bibnamefont {Sank}}, \bibinfo {author} {\bibfnamefont {H.}~\bibnamefont
  {Wang}}, \bibinfo {author} {\bibfnamefont {J.}~\bibnamefont {Wenner}},
  \bibinfo {author} {\bibfnamefont {A.~N.}\ \bibnamefont {Cleland}}, \bibinfo
  {author} {\bibfnamefont {M.~R.}\ \bibnamefont {Geller}}, \ and\ \bibinfo
  {author} {\bibfnamefont {J.~M.}\ \bibnamefont {Martinis}},\ }\href {\doibase
  10.1126/science.1173440} {\bibfield  {journal} {\bibinfo  {journal}
  {Science}\ }\textbf {\bibinfo {volume} {325}},\ \bibinfo {pages} {722}
  (\bibinfo {year} {2009})}\BibitemShut {NoStop}%
\bibitem [{\citenamefont {Barends}\ \emph {et~al.}(2014)\citenamefont
  {Barends}, \citenamefont {Kelly}, \citenamefont {Megrant}, \citenamefont
  {Veitia}, \citenamefont {Sank}, \citenamefont {Jeffrey}, \citenamefont
  {White}, \citenamefont {Mutus}, \citenamefont {Fowler}, \citenamefont
  {Campbell}, \citenamefont {Chen}, \citenamefont {Chen}, \citenamefont
  {Chiaro}, \citenamefont {Dunsworth}, \citenamefont {Neill}, \citenamefont
  {O'Malley}, \citenamefont {Roushan}, \citenamefont {Vainsencher},
  \citenamefont {Wenner}, \citenamefont {Korotkov}, \citenamefont {Cleland},\
  and\ \citenamefont {Martinis}}]{Barends2014}%
  \BibitemOpen
  \bibfield  {author} {\bibinfo {author} {\bibfnamefont {R.}~\bibnamefont
  {Barends}}, \bibinfo {author} {\bibfnamefont {J.}~\bibnamefont {Kelly}},
  \bibinfo {author} {\bibfnamefont {A.}~\bibnamefont {Megrant}}, \bibinfo
  {author} {\bibfnamefont {A.}~\bibnamefont {Veitia}}, \bibinfo {author}
  {\bibfnamefont {D.}~\bibnamefont {Sank}}, \bibinfo {author} {\bibfnamefont
  {E.}~\bibnamefont {Jeffrey}}, \bibinfo {author} {\bibfnamefont {T.~C.}\
  \bibnamefont {White}}, \bibinfo {author} {\bibfnamefont {J.}~\bibnamefont
  {Mutus}}, \bibinfo {author} {\bibfnamefont {A.~G.}\ \bibnamefont {Fowler}},
  \bibinfo {author} {\bibfnamefont {B.}~\bibnamefont {Campbell}}, \bibinfo
  {author} {\bibfnamefont {Y.}~\bibnamefont {Chen}}, \bibinfo {author}
  {\bibfnamefont {Z.}~\bibnamefont {Chen}}, \bibinfo {author} {\bibfnamefont
  {B.}~\bibnamefont {Chiaro}}, \bibinfo {author} {\bibfnamefont
  {A.}~\bibnamefont {Dunsworth}}, \bibinfo {author} {\bibfnamefont
  {C.}~\bibnamefont {Neill}}, \bibinfo {author} {\bibfnamefont
  {P.}~\bibnamefont {O'Malley}}, \bibinfo {author} {\bibfnamefont
  {P.}~\bibnamefont {Roushan}}, \bibinfo {author} {\bibfnamefont
  {A.}~\bibnamefont {Vainsencher}}, \bibinfo {author} {\bibfnamefont
  {J.}~\bibnamefont {Wenner}}, \bibinfo {author} {\bibfnamefont {A.~N.}\
  \bibnamefont {Korotkov}}, \bibinfo {author} {\bibfnamefont {A.~N.}\
  \bibnamefont {Cleland}}, \ and\ \bibinfo {author} {\bibfnamefont {J.~M.}\
  \bibnamefont {Martinis}},\ }\href {http://dx.doi.org/10.1038/nature13171}
  {\bibfield  {journal} {\bibinfo  {journal} {Nature}\ }\textbf {\bibinfo
  {volume} {508}},\ \bibinfo {pages} {500} (\bibinfo {year}
  {2014})}\BibitemShut {NoStop}%
\bibitem [{\citenamefont {Chen}\ \emph {et~al.}(2014)\citenamefont {Chen},
  \citenamefont {Neill}, \citenamefont {Roushan}, \citenamefont {Leung},
  \citenamefont {Fang}, \citenamefont {Barends}, \citenamefont {Kelly},
  \citenamefont {Campbell}, \citenamefont {Chen}, \citenamefont {Chiaro},
  \citenamefont {Dunsworth}, \citenamefont {Jeffrey}, \citenamefont {Megrant},
  \citenamefont {Mutus}, \citenamefont {O'Malley}, \citenamefont {Quintana},
  \citenamefont {Sank}, \citenamefont {Vainsencher}, \citenamefont {Wenner},
  \citenamefont {White}, \citenamefont {Geller}, \citenamefont {Cleland},\ and\
  \citenamefont {Martinis}}]{chen2014qubit}%
  \BibitemOpen
  \bibfield  {author} {\bibinfo {author} {\bibfnamefont {Y.}~\bibnamefont
  {Chen}}, \bibinfo {author} {\bibfnamefont {C.}~\bibnamefont {Neill}},
  \bibinfo {author} {\bibfnamefont {P.}~\bibnamefont {Roushan}}, \bibinfo
  {author} {\bibfnamefont {N.}~\bibnamefont {Leung}}, \bibinfo {author}
  {\bibfnamefont {M.}~\bibnamefont {Fang}}, \bibinfo {author} {\bibfnamefont
  {R.}~\bibnamefont {Barends}}, \bibinfo {author} {\bibfnamefont
  {J.}~\bibnamefont {Kelly}}, \bibinfo {author} {\bibfnamefont
  {B.}~\bibnamefont {Campbell}}, \bibinfo {author} {\bibfnamefont
  {Z.}~\bibnamefont {Chen}}, \bibinfo {author} {\bibfnamefont {B.}~\bibnamefont
  {Chiaro}}, \bibinfo {author} {\bibfnamefont {A.}~\bibnamefont {Dunsworth}},
  \bibinfo {author} {\bibfnamefont {E.}~\bibnamefont {Jeffrey}}, \bibinfo
  {author} {\bibfnamefont {A.}~\bibnamefont {Megrant}}, \bibinfo {author}
  {\bibfnamefont {J.~Y.}\ \bibnamefont {Mutus}}, \bibinfo {author}
  {\bibfnamefont {P.~J.~J.}\ \bibnamefont {O'Malley}}, \bibinfo {author}
  {\bibfnamefont {C.~M.}\ \bibnamefont {Quintana}}, \bibinfo {author}
  {\bibfnamefont {D.}~\bibnamefont {Sank}}, \bibinfo {author} {\bibfnamefont
  {A.}~\bibnamefont {Vainsencher}}, \bibinfo {author} {\bibfnamefont
  {J.}~\bibnamefont {Wenner}}, \bibinfo {author} {\bibfnamefont {T.~C.}\
  \bibnamefont {White}}, \bibinfo {author} {\bibfnamefont {M.~R.}\ \bibnamefont
  {Geller}}, \bibinfo {author} {\bibfnamefont {A.~N.}\ \bibnamefont {Cleland}},
  \ and\ \bibinfo {author} {\bibfnamefont {J.~M.}\ \bibnamefont {Martinis}},\
  }\href {\doibase 10.1103/PhysRevLett.113.220502} {\bibfield  {journal}
  {\bibinfo  {journal} {Phys. Rev. Lett.}\ }\textbf {\bibinfo {volume} {113}},\
  \bibinfo {pages} {220502} (\bibinfo {year} {2014})}\BibitemShut {NoStop}%
\bibitem [{\citenamefont {{Geller}}\ \emph {et~al.}(2014)\citenamefont
  {{Geller}}, \citenamefont {{Donate}}, \citenamefont {{Chen}}, \citenamefont
  {{Neill}}, \citenamefont {{Roushan}},\ and\ \citenamefont
  {{Martinis}}}]{geller2014tunable}%
  \BibitemOpen
  \bibfield  {author} {\bibinfo {author} {\bibfnamefont {M.~R.}\ \bibnamefont
  {{Geller}}}, \bibinfo {author} {\bibfnamefont {E.}~\bibnamefont {{Donate}}},
  \bibinfo {author} {\bibfnamefont {Y.}~\bibnamefont {{Chen}}}, \bibinfo
  {author} {\bibfnamefont {C.}~\bibnamefont {{Neill}}}, \bibinfo {author}
  {\bibfnamefont {P.}~\bibnamefont {{Roushan}}}, \ and\ \bibinfo {author}
  {\bibfnamefont {J.~M.}\ \bibnamefont {{Martinis}}},\ }\href@noop {}
  {\bibfield  {journal} {\bibinfo  {journal} {ArXiv e-prints}\ } (\bibinfo
  {year} {2014})},\ \Eprint {http://arxiv.org/abs/1405.1915} {arXiv:1405.1915
  [quant-ph]} \BibitemShut {NoStop}%
\bibitem [{\citenamefont {Ghosh}\ \emph
  {et~al.}(2013{\natexlab{a}})\citenamefont {Ghosh}, \citenamefont
  {Galiautdinov}, \citenamefont {Zhou}, \citenamefont {Korotkov}, \citenamefont
  {Martinis},\ and\ \citenamefont {Geller}}]{PhysRevA.87.022309}%
  \BibitemOpen
  \bibfield  {author} {\bibinfo {author} {\bibfnamefont {J.}~\bibnamefont
  {Ghosh}}, \bibinfo {author} {\bibfnamefont {A.}~\bibnamefont {Galiautdinov}},
  \bibinfo {author} {\bibfnamefont {Z.}~\bibnamefont {Zhou}}, \bibinfo {author}
  {\bibfnamefont {A.~N.}\ \bibnamefont {Korotkov}}, \bibinfo {author}
  {\bibfnamefont {J.~M.}\ \bibnamefont {Martinis}}, \ and\ \bibinfo {author}
  {\bibfnamefont {M.~R.}\ \bibnamefont {Geller}},\ }\href {\doibase
  10.1103/PhysRevA.87.022309} {\bibfield  {journal} {\bibinfo  {journal} {Phys.
  Rev. A}\ }\textbf {\bibinfo {volume} {87}},\ \bibinfo {pages} {022309}
  (\bibinfo {year} {2013}{\natexlab{a}})}\BibitemShut {NoStop}%
\bibitem [{Note1()}]{Note1}%
  \BibitemOpen
  \bibinfo {note} {One can construct an arbitrary pulse shape that satisfies
  our Eq.(\ref {eq:SWAPqutrit}). However, we here analyze trapezoidal pulse as
  it is analytically tractable as well as closely approximates a realistic
  pulse generated by the control electronics for superconducting
  circuits}\BibitemShut {NoStop}%
\bibitem [{\citenamefont {Ghosh}\ \emph
  {et~al.}(2013{\natexlab{b}})\citenamefont {Ghosh}, \citenamefont {Fowler},
  \citenamefont {Martinis},\ and\ \citenamefont {Geller}}]{PhysRevA.88.062329}%
  \BibitemOpen
  \bibfield  {author} {\bibinfo {author} {\bibfnamefont {J.}~\bibnamefont
  {Ghosh}}, \bibinfo {author} {\bibfnamefont {A.~G.}\ \bibnamefont {Fowler}},
  \bibinfo {author} {\bibfnamefont {J.~M.}\ \bibnamefont {Martinis}}, \ and\
  \bibinfo {author} {\bibfnamefont {M.~R.}\ \bibnamefont {Geller}},\ }\href
  {\doibase 10.1103/PhysRevA.88.062329} {\bibfield  {journal} {\bibinfo
  {journal} {Phys. Rev. A}\ }\textbf {\bibinfo {volume} {88}},\ \bibinfo
  {pages} {062329} (\bibinfo {year} {2013}{\natexlab{b}})}\BibitemShut
  {NoStop}%
\bibitem [{\citenamefont {Ronke}\ \emph {et~al.}(2011)\citenamefont {Ronke},
  \citenamefont {Spiller},\ and\ \citenamefont {D'Amico}}]{PhysRevA.83.012325}%
  \BibitemOpen
  \bibfield  {author} {\bibinfo {author} {\bibfnamefont {R.}~\bibnamefont
  {Ronke}}, \bibinfo {author} {\bibfnamefont {T.~P.}\ \bibnamefont {Spiller}},
  \ and\ \bibinfo {author} {\bibfnamefont {I.}~\bibnamefont {D'Amico}},\ }\href
  {\doibase 10.1103/PhysRevA.83.012325} {\bibfield  {journal} {\bibinfo
  {journal} {Phys. Rev. A}\ }\textbf {\bibinfo {volume} {83}},\ \bibinfo
  {pages} {012325} (\bibinfo {year} {2011})}\BibitemShut {NoStop}%
\bibitem [{\citenamefont {Liu}\ \emph {et~al.}(2004)\citenamefont {Liu},
  \citenamefont {\"Ozdemir}, \citenamefont {Miranowicz},\ and\ \citenamefont
  {Imoto}}]{PhysRevA.70.042308}%
  \BibitemOpen
  \bibfield  {author} {\bibinfo {author} {\bibfnamefont {Y.-x.}\ \bibnamefont
  {Liu}}, \bibinfo {author} {\bibfnamefont {i.~m. c.~K.}\ \bibnamefont
  {\"Ozdemir}}, \bibinfo {author} {\bibfnamefont {A.}~\bibnamefont
  {Miranowicz}}, \ and\ \bibinfo {author} {\bibfnamefont {N.}~\bibnamefont
  {Imoto}},\ }\href {\doibase 10.1103/PhysRevA.70.042308} {\bibfield  {journal}
  {\bibinfo  {journal} {Phys. Rev. A}\ }\textbf {\bibinfo {volume} {70}},\
  \bibinfo {pages} {042308} (\bibinfo {year} {2004})}\BibitemShut {NoStop}%
\end{thebibliography}%

\end{document}